\documentclass[twocolumn,amsmath,amssymb,nofootinbib,eqsecnum,tighten,prd]{revtex4}

\usepackage{fancybox}
\usepackage{graphicx}
\usepackage{dcolumn} 
\usepackage{bm}      
\usepackage{curves}
\usepackage{color}

 \def\ep{{\epsilon}}

 \def\frac#1#2{{#1\over #2}}

 \def\s{\sqrt}

 \def\al{\alpha'}
 \def\de{\partial}

 \def\f {\frac}

 \def\ddd{\cdot\cdot\cdot}

 \def\ep{\epsilon}

 \def\vp{\varphi}

 \def\ep{{\epsilon}}

 \def\frac#1#2{{#1\over #2}}

 \def\s{\sqrt}

\def\be{\begin{equation}}
\def\ee{\end{equation}}
\def\ba{\begin{eqnarray}}
\def\ea{\end{eqnarray}}

\begin{document}

\title{Holographic Derivation of
Entanglement Entropy from AdS/CFT}

\author{Shinsei Ryu}
\affiliation{Kavli Institute for Theoretical Physics,
         University of California,
         Santa Barbara,
         CA 93106,
         USA}
\author{Tadashi Takayanagi}
\affiliation{Kavli Institute for Theoretical Physics,
         University of California,
         Santa Barbara,
         CA 93106,
         USA}
\date{\today}

\begin{abstract}
A holographic derivation of the entanglement entropy in quantum
(conformal) field theories is proposed from AdS/CFT correspondence.
We argue that the entanglement entropy in $d+1$ dimensional
conformal field theories can be obtained from the area of $d$
dimensional minimal surfaces in AdS$_{d+2}$, analogous to the
Bekenstein-Hawking formula for black hole entropy. We show that our
proposal perfectly reproduces the correct entanglement entropy in 2D
CFT when applied to AdS$_3$. We also compare the entropy computed in
AdS$_5\times$S$^5$ with that of the free ${\mathcal N}=4$ super
Yang-Mills.
\end{abstract}

\maketitle

\section{Introduction}
\setcounter{section}{1}

One of the most remarkable successes in gravitational aspects of
string theory is the microscopic derivation of the
Bekenstein-Hawking entropy $S_{BH}$
\begin{equation}
S_{BH}=\frac{{\rm Area~of~horizon}}{4G_N},
\label{BHentropy}
\end{equation}
for BPS black holes \cite{StVa}. This idea relates the gravitational
entropy with the degeneracy of quantum field theory as its
microscopic description. Taking near horizon limit, we can regard
this as a special example of AdS/CFT correspondence
\cite{Maldacena,ADSGKP,ADSWitten}. It claims that the $d+1$
dimensional conformal field theories (CFT$_{d+1}$) are equivalent to
the (super)gravity on $d+2$ dimensional anti-deSitter space
AdS$_{d+2}$. We expect that each CFT is sitting at the boundary of
AdS space.

On the other hand, there is a different kind of entropy called
entanglement entropy (von-Neumann entropy) in quantum mechanical
systems. The entanglement entropy
\begin{eqnarray}
S_A = - \mathrm{tr}_{A}\, \rho_{A} \log \rho_{A},\quad
\rho_{A}=\mathrm{tr}_{B}\, |\Psi\rangle \langle \Psi|, \label{eq:
def entanglement entropy}
\end{eqnarray}
provides us with a convenient way to measure how closely entangled
(or how ``quantum'') a given wave function $|\Psi\rangle$ is. Here,
the total system is divided into two subsystems $A$ and $B$ and
$\rho_{A}$ is the reduced density matrix for the subsystem $A$
obtained by taking a partial trace over the subsystem $B$ of the
total density matrix $\rho=|\Psi\rangle \langle \Psi|$. Intuitively,
we can think $S_A$ as the entropy for an observer who is only
accessible to the subsystem $A$ and cannot receive any signals from
$B$. In this sense, the subsystem $B$ is analogous to the inside of
a black hole horizon for an observer sitting in $A$, i.e., outside
of the horizon. Indeed, an original motivation of the entanglement
entropy was its similarity to the Bekenstein-Hawking entropy
\cite{Bombelli,Srednicki}.

The entanglement entropy is of growing importance in many fields of
physics in our exploration for better understanding of quantum
systems. For example, in a modern trend of condensed matter physics
it has been becoming clear that quantum phases of matter need to be
characterized by their pattern of entanglement encoded in many-body
wave functions of ground states, rather than conventional order
parameters \cite{VLRK,Kitaev05, Levin05}. Recently, the entanglement
entropy has been extensively studied in low-dimensional quantum
many-body systems as a new tool to investigate the nature of quantum
criticality (refer to \cite{Calabrese04} and references therein for
example).

For one-dimensional (1D) quantum many-body systems at criticality
(i.e. 2D CFT), it is known that the entanglement entropy is given by
\cite{Holzhey94,Calabrese04}
\begin{eqnarray}
S_A&=&\frac{c}{3}\cdot\log\left(\f{L}{\pi a}\sin\left(\f{\pi
l}{L}\right)\right),
 \label{eq: EE for CFT_2}
\end{eqnarray}
where $l$ and $L$ are the length of the subsystem $A$ and the total
system $A\cup B$ (both ends of $A\cup B$ are periodically
identified), respectively; $a$ is a ultra violet (UV) cutoff
(lattice spacing); $c$ is the central charge of the CFT. When we are
away from criticality, Eq.\ (\ref{eq: EE for CFT_2}) is replaced by
\cite{VLRK,Calabrese04}
\begin{eqnarray}
S_A &=&
\frac{c}{6}\cdot{\cal A}\cdot\log \f{\xi}{a},
\label{eq: EE for massive}
\end{eqnarray}
where $\xi$ is the correlation length and $\cal A$ is the number
of boundary points of $A$
(e.g. ${\cal A}=2$ in the setup of  (\ref{eq: EE for CFT_2})).

In spite of
these recent developments, and
its similarity to the black hole entropy,
a comprehensive gravitational interpretation of the entanglement
entropy has been lacking so far.  Here,  we present a simple proposal
on this issue in the light of AdS/CFT duality.
Earlier discussions from
different viewpoints can be
found in e.g.  papers \ \cite{HMS,MBH}. Define
the entanglement entropy $S_A$ in
a CFT on
$\mathbb{R}^{1,d}$ (or $\mathbb{R}\times S^{d}$)
for a subsystem $A$ that has an arbitrary $d-1$ dimensional
boundary $\de A\in \mathbb{R}^{d}$ $({\rm or}~ S^{d})$.
In this setup we propose the following `area law'
\begin{equation} S_{A}=\frac{{\rm Area~of}~\gamma_{A}}{4G^{(d+2)}_N},
\label{arealaw}
\end{equation}
where $\gamma_{A}$ is the $d$ dimensional static minimal surface in
AdS$_{d+2}$ whose boundary is given by $\de A$, and $G^{(d+2)}_{N}$
is the $d+2$ dimensional Newton constant. Intuitively, this suggests
that the minimal surface $\gamma_A$ plays the role of a holographic
screen for an observer who is only accessible to the subsystem $A$.
We show explicitly the relation (\ref{arealaw}) in the lowest
dimensional case $d=1$, where $\gamma_{A}$ is given by a geodesic
line in AdS$_3$. We also compute $S_A$ from the gravity side for
general $d$ and compare it with field theory results, which is
successful at least qualitatively. From (\ref{arealaw}), it is
readily seen that the basic properties of the entanglement entropy
(i) $S_A=S_B$ ($B$ is the complement of $A$) and (ii)
$S_{A_1}+S_{A_2}\geq S_{A_1\cup A_2}$ (subadditivity) are satisfied.

We can also define the entanglement entropy at finite temperature $T=\beta^{-1}$.
E.g. in a 2D CFT on a infinitely long line, it is given by
replacing $L$ in Eq.\ (\ref{eq: EE for CFT_2})
with $i\beta$ \cite{Calabrese04}.   We argue that Eq.\ (\ref{arealaw}) still holds
in $T> 0$ cases. Note that $S_A=S_B$ is no longer true if $T> 0$ since
$\rho$ is in a mixed state generically. At high temperature,
we will see that this occurs due to the presence of black hole horizon
in the dual gravity description.

\section{Entanglement Entropy in $\mathrm{AdS}_3/\mathrm{CFT}_2$}

\begin{figure}
\begin{center}
\includegraphics[width=8cm,clip]{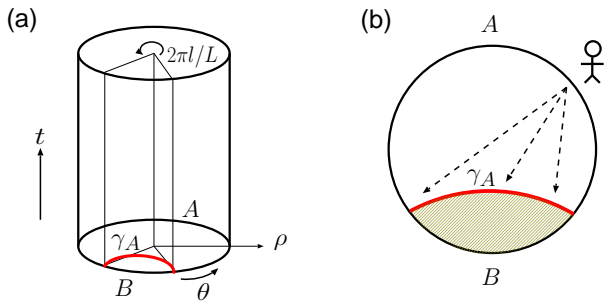}
\end{center}
\caption{
\label{fig: ads3_cft2.eps}
(a)
AdS$_3$ space and CFT$_2$ living on its boundary
and (b) a geodesics $\gamma_A$ as a
holographic screen.
}
\end{figure}

Let us start with the entanglement entropy in 2D CFTs.
According to AdS/CFT correspondence, gravitational theories on
AdS$_3$ space of radius $R$ are dual to (1+1)D CFTs with
the central charge  \cite{BH}
\be c=\frac{3R}{2G^{(3)}_N}. \label{centralcharge} \ee
The metric of AdS$_3$ in the
global coordinate $(t,\rho,\theta)$  is
\begin{eqnarray}
ds^2=R^2\left(-\cosh\rho^2
dt^2+d\rho^2+\sinh\rho^2d\theta^2\right). \label{ads}
\end{eqnarray}
At the boundary $\rho=\infty$ of AdS$_3$ the metric is
divergent. To regulate physical quantities we put a
cutoff $\rho_0$ and restrict the space to the bounded region
$\rho\leq\rho_0$. This procedure corresponds to
the UV cutoff in the dual CFTs \cite{SuWi}.
If $L$ is the total
length of the system with both ends identified, and $a$ is the
lattice spacing (or UV cutoff) in the CFTs,
we have the relation (up to a numerical factor)
\begin{eqnarray}
e^{\rho_0}\sim L/a. \label{cutoff}
\label{delti}
\end{eqnarray}

The (1+1)D spacetime for the CFT$_2$ is identified with the cylinder
$(t,\theta)$ at the boundary $\rho=\rho_0$. The subsystem $A$ is the
region $0\leq \theta\leq 2\pi l/L$. Then $\gamma_A$ in  Eq.\
(\ref{arealaw}) is identified with the static geodesic
 that connects the boundary points
$\theta=0$ and $2\pi l/L$ with $t$ fixed,
traveling inside AdS$_3$
(Fig.\ \ref{fig: ads3_cft2.eps} (a)).
With the cutoff $\rho_0$ introduced above,
the geodesic distance $L_{\gamma_A}$
is given by
\begin{eqnarray}
\cosh\left(\f{L_{\gamma_A}}{R}\right)=1+2\sinh^2\rho_0~\sin^2\frac{\pi l}{L}.
\label{geodesic}
\end{eqnarray}

Assuming the large UV cutoff  $e^{\rho_0}\gg 1$,
the entropy (\ref{arealaw}) is expressed as follows, using Eq.\ (\ref{centralcharge})
\begin{eqnarray}
S_A
\!\simeq\!
\frac{R}{4G^{(3)}_N}\!\log\left(\!e^{2\rho_0}\sin^2\frac{\pi l}{L}\right)
\! =\!
\frac{c}{3}\!\log\left(\!e^{\rho_0}\sin\frac{\pi l}{L}\right).~~
\end{eqnarray}
This entropy precisely coincides with the known CFT result
(\ref{eq: EE for CFT_2})
after we remember the relation Eq.\ (\ref{delti}).

This proposed relation (\ref{arealaw}) suggests that the geodesic
 (or minimal surface in the higher dimensional case)
 $\gamma_A$ is analogous to an event
 horizon as if $B$ were a black hole, though
 the division into $A$ and $B$ is just artificial. In other
words, the observer, who is not accessible to $B$, will probe
$\gamma_A$ as a holographic screen \cite{holography}, instead of $B$
itself (Fig.\ \ref{fig: ads3_cft2.eps} (b)). The minimal surface
provides the severest entropy bound when we fix its boundary
condition. In our case it saturates the bound.

More generally,  we can consider a subsystem $A$ which consists of
multiple disjoint intervals as follows
 \be A=\{x|x\in [r_1,s_1]\cup [r_2,s_2]\cup \ddd \cup
[r_{N},s_{N}]\}, \label{region}
\ee
where $0\leq r_1<s_1<r_2< s_2<
\ddd < r_N<s_N\leq L$.  In the dual AdS$_3$ description,
the region (\ref{region})
corresponds to $\theta\in \cup_{i=1}^N[\frac{2\pi
r_i}{L},\frac{2\pi s_i}{L}]$ at the boundary. In this case it is not
straightforward to determine minimal (geodesic) lines responsible for the entropy.
However, we can find the answer from the entanglement entropy
computed in the CFT side.
 The general prescription of calculating
the entropy for such systems
is
given in \cite{Calabrese04}
using conformal mapping.
For our system (\ref{region}),
we find,
when rewritten in the AdS$_3$ language,
 the following expression of $S_A$
\begin{eqnarray}
S_A\!&=&\!\frac{
\sum_{i,j}L_{r_j,s_i}\!
-\sum_{i<j}L_{r_j,r_i}\!
-\sum_{i<j}L_{s_j,s_i}}{4G^{(3)}_N},~~ \label{multi}
\end{eqnarray}
where $L_{a,b}$ is the geodesic distance
between two boundary points $a$ and $b$.
We can think that the correct definition of minimal surface is given by the numerator
in Eq.\ (\ref{multi}).

Next we turn to the entanglement entropy at finite temperature.
We assume the spacial length of the total system $L$ is
infinitely long s.t. $\beta/L \ll 1$.
At high temperature, the gravity dual of the CFT is
the Euclidean BTZ black hole \cite{BTZ}
with the metric given by
\be
ds^2=(r^2-r_+^2)d\tau^2+ \f{R^2}{r^2-r^2_{+}}dr^2+r^2
d\vp^2. \label{BTZ}
\ee
The Euclidean time is compactified as $\tau\sim\tau+\f{2\pi
R}{r_+}$ to obtain a smooth geometry in addition to the periodicity
$\vp\sim \vp+2\pi$. Looking at its boundary, we
find the relation $\f{\beta}{L}=\f{R}{r_{+}}\ll 1$
between the CFT and the BTZ black hole.

The subsystem $A$ is defined by $0\leq \vp\leq 2\pi l/L$ at the
boundary. Then we expect
that the entropy can be computed from the geodesic distance
between the boundary points $\vp=0, 2\pi l/L$  at a fixed time. To find
the geodesic line, it is useful to remember the familiar fact that the
Euclidean BTZ black hole at temperature $T_{BTZ}$ is equivalent to the thermal
AdS$_3$ at temperature $1/T_{BTZ}$. This equivalence can be interpreted as
a modular transformation in the CFT side \cite{MS}. If we define the new
coordinates $r=r_{+}\cosh\rho,\
r_+\tau=R\theta,\  r_+\vp=Rt, $
then the metric  (\ref{BTZ})  indeed becomes the Euclidean version of AdS$_3$  metric
(\ref{ads}).
Thus the geodesic distance can be found in the same way
as in Eq.\ (\ref{geodesic}) : $\cosh(L_{\gamma_A}/R)
=1+2\cosh^2\rho_0\sinh^2\left(\f{\pi l}{\beta}\right)$,
where the UV cutoff is interpreted as $e^{\rho_0}\sim \beta/a$.
Then the area law
(\ref{arealaw}) reproduces the known CFT result \cite{Calabrese04}
\be
S_{A}(\beta)=\f{c}{3}\cdot\log\left(\f{\beta}{\pi a}\sinh\left(\f{\pi l}{
 \beta}\right)\right). \ee
We can extend these arguments to the multi interval cases and find the same
formula (\ref{multi}) as before.

It is instructive to repeat the same analysis in the Poincare
metric $ ds^2=R^2 z^{-2}(dz^2-dx_0^2+dr^2). $
We define the subsystem $A$ by the region $-l/2\leq r\leq l/2$ at the boundary  $z=0$.
The geodesic line $\gamma_A$ is given by  \be (r,z)=\f{l}{2}(\cos
s,\sin s),\ \ \ \ (\ep\leq s \leq \pi-\ep), \label{minc} \ee where
the infinitesimal $\ep$ is
the UV cutoff $\ep\sim 2a/l$ (or equally $z_{UV}\sim a$). We obtain the entropy $S_A$
as follows \be S_A=\f{L_{\gamma_A}}
{4G^{(3)}_N}=\f{R}{2G^{(3)}_N}\int^{\pi/2}_\ep
\f{ds}{\sin s}=\f{c}{3}\log\f{l}{a}. \ee
This reproduces  the small $l$ limit
of Eq. (\ref{eq: EE for CFT_2}) \cite{Holzhey94} .

When we perturb a CFT by a relevant perturbation, the RG flow
generically drives the theory to a trivial IR fixed point.  We
denote the correlation length $\xi$ in the latter theory. In the
AdS dual, this massive deformation corresponds to capping
off the IR region, restricting the allowed values of $z$ to $z\leq
\xi$. In the large $l$ limit, we find \be
S_A =\f{1}{4G^{(3)}_N} \int^{2\xi/l}_{\ep}\f{ds}{\sin
s}=\f{c}{6}\log\f{\xi}{a}. \ee
This agrees with the CFT result
with $\mathcal{A}=1$
(\ref{eq: EE for massive})  \cite{VLRK,Calabrese04}.

\section{Higher Dimensional Cases}

Motivated by the success in
 our gravitational derivation of the entanglement
entropy for $d=1$, it is interesting to extend the idea to higher dimensional
cases ($d\geq 2$).
A natural thing to do is to replace
geodesic lines with minimal surfaces.
The computations are
analogous to the evaluation
of Wilson loops \cite{Wilsonline}, though the dimension of relevant
minimal surfaces is different.

We will work in the Poincare metric for AdS$_{d+2}$ \be ds^2=R^2
z^{-2} (dz^2-dx_0^2+\sum_{i=1}^d dx_i^2).  \label{Poincare} \ee We
consider two examples for the shape of $A$. The first one is a
straight belt $A_{S}=\{x_i|x_1\in [-l/2,l/2], x_{2,3,\ddd,d}\in
[-\infty,\infty]\}$ at the boundary $z=0$ (Fig.\ 2 (a)). In this
case the minimal surface is defined by $
dz/dx_1=\s{z^{2d}_*-z^{2d}}/z^d, $ where $z_*$ is determined by $
l/2=\int^{z_*}_{0}dz z^d (z^{2d}_*-z^{2d})^{-1/2} =z_* \s{\pi}
\Gamma(\f{d+1}{2d})/\Gamma(\f{1}{2d}). $ The area of this minimal
surface is \be {\rm
Area}_{A_{S}}\!=\!\f{2R^d}{d-1}\!\!\left(\f{L}{a}\right)^{d-1}\!\!\!\!
-\f{2^d \pi^{d/2} R^{d}}{d-1}\!\!\left(\f{\Gamma(\f{d+1}{2d})}
{\Gamma(\f{1}{2d})}\right)^{\!\!\!d}\!\!\!
\left(\f{L}{l}\right)^{\!\!\! d-1}\!\!\!\!\!\!\!, \label{areaone}
\ee where $L$ is the length of $A_S$ in the
$x_{2,3,\cdots,d}$-direction.

The second example
is the disk $A_{D}$ defined by $A_{D}=\{x_i|r\leq l\}$ (Fig.\ 2(b))
in the polar coordinate $\sum_{i}dx_i^2=dr^2+r^2d\Omega_{d-1}^2$.
The minimal surface is a $d$ dimensional ball $B^d$ defined by
(\ref{minc}). Its area is
\begin{eqnarray}
&&{\rm Area}_{A_{D}}
=
C
\int^1_{a/l}dy \f{(1-y^2)^{(d-2)/2}}{y^d}  \nonumber \\
&&=
  p_1  \left(l/a\right)^{d-1}
+ p_3 \left(l/a \right)^{d-3}
+\cdots  \label{areatwo}  \\
&&
\cdots
+\left\{
\begin{array}{ll}
\displaystyle
p_{d-1}\left(l/a\right)
+
p_d
+
O(a/l), &
 \mbox{$d$: even},   \\
\displaystyle
p_{d-2} \left(l/a\right)^{2}
+ q \log \left(l/a\right)+ O(1),
&  \mbox{$d$: odd,}   \\
\end{array}
\right.
 \nonumber
\end{eqnarray}
where $C=2\pi^{d/2}R^d/ \Gamma(d/2)$ and
$p_1/C = (d-1)^{-1}$ etc.
For $d$ even,
$p_d/C = (2\s{\pi})^{-1}\Gamma\left(\f{d}{2}\right)
\Gamma\left(\f{1-d}{2}\right)$
and for $d$ odd,
$q/C = (-)^{(d-1)/2}(d-2)!!/(d-1)!!$.

\begin{figure}
\begin{center}
\includegraphics[width=8cm,clip]{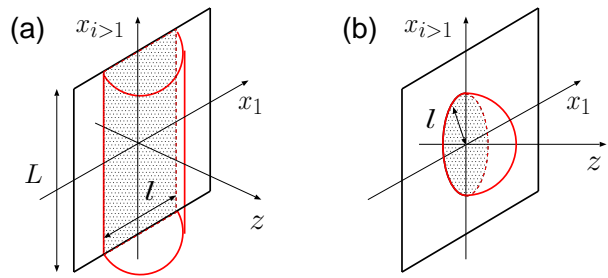}
\end{center}
\caption{
\label{fig: min_surf}
Minimal surfaces in
AdS$_{d+2}$: (a) $A_S$
and (b) $A_D$.
}
\end{figure}

{}From these results, the entanglement entropy can be calculated by
Eq.\ (\ref{arealaw}). Each of (\ref{areaone}) and (\ref{areatwo})
has a UV divergent term $\sim a^{-d+1}$ that is proportional to the
area of the boundary $\de A$. This agrees with the known `area law'
of the entanglement entropy in quantum field theories
\cite{Bombelli,Srednicki}. Note that this `area law' is related to
ours Eq.\ (\ref{arealaw})  via the basic property of holography.

We may prefer a physical quantity that is independent of the cutoff
(i.e.~universal). The second term in Eq.\ (\ref{areaone}) has this
property. In general, when $A$ is a finite size, there is a
universal and conformal invariant constant contribution to $S_A$ if
$d$ is even (see \cite{GrWi} for properties of minimal surfaces in
AdS). In (2+1)D topological field theories the constant contribution
to $S_A$ encodes the quantum dimension and is called the topological
entanglement entropy \cite{Kitaev05, Levin05}.  If $d$ is odd, the
coefficient of the logarithmic term $\sim \log(l/a)$ is universal as
in Eq.\ (\ref{eq: EE for CFT_2}).

Let us apply the previous results to a specific string theory setup.
Type IIB string on AdS$_5\times S^5$ is dual to 4D $\mathcal{N}=4$
$SU(N)$ super Yang-Mills theory \cite{Maldacena}. 
The radius of AdS$_5$ and $S^5$ are
given by the same value $R=(4\pi g_s \al^2 N)^{\f{1}{4}}$. The 5D Newton constant
is related to the 10D one via $G^{(10)}_N=\pi^3R^5 G^{(5)}_N $. Then
we obtain from Eqs.\ (\ref{areaone}) and (\ref{areatwo}) \ba
S_{A_{S}}&=&\f{N^2L^2}{2\pi a^2}-2\s{\pi}
\left(\f{\Gamma\left(\f23\right)}{\Gamma\left(\f16\right)}
\right)^3\f{N^2L^2}{l^2}, \label{sta}\\
S_{A_{D}}&=&
N^2\left[\f{l^2}{a^2}-\log\left(\f{l}{a}\right)+O(1)\right].
\ea

It is interesting to compare the finite universal term in Eq.\
(\ref{sta}) with the field theory one. For free real scalars and
 fermions in general dimensions, one way to compute $S_{A_S}$
is presented in \cite{Casini} (see also \cite{Fursaev}). Indeed, this leads to the same
behavior in $a$ and $l$ as in Eq.\ (\ref{sta}). Following this
approach, we can estimate finite contributions from 6 scalars and 4
Majorana fermions in the $\mathcal{N}=4$ Yang-Mills multiplet.  In
the end, we obtain $S^{freeCFT}_{finite}\sim -(0.068+g)\cdot
N^2L^2/l^2$, where $g$ is the contribution from the gauge field
($g=0.010$ if we treat the gauge field as 2 scalars). On the other
hand, our AdS$_5$ result (\ref{sta}) leads to $S^{AdS}_{finite}\sim
-0.051\cdot N^2L^2/l^2$. We may think this is a good agreement if we
remember that the gravity description corresponds to the strongly
coupled gauge theory
 instead of the free theory as in \cite{GKP}.

We can also compute the entanglement entropy for the near horizon limit
AdS$_{4}\times S^{7}$ (AdS$_{7}\times S^{4}$) of $N$ $M2$-branes
($M5$-branes)
\ba
S^{M2}_{A_S}&=&\f{\s{2}}{3}N^{3/2}\left[\f{L}{a}-
\f{4\pi^3}{\Gamma(1/4)^4}\f{L}{l}\right], \label{sma} \\
S^{M2}_{A_D}&=&\f{\s{2}\pi}{3}N^{3/2}\left[\f{l}{a}-1\right],  \label{smb}\\
 S^{M5}_{A_S}&=&\!\f{2}{3\pi^2}N^{3}\!\left[\f{L^4}{a^4}-
16\pi^{5/2}\!\f{\Gamma(3/5)^5}{\Gamma(1/10)^5}\f{L^4}{l^4}\right]\!,  \label{smc}\\
S^{M5}_{A_D}&=&\f{32}{9}N^{3}\left[\f{1}{4}\cdot\f{l^4}{a^4}
-\f{3}{4}\cdot\f{l^2}{a^2}+\f{3}{8}\log\f{l}{a}\right]. \ea Note
that the constant terms in Eqs.\ (\ref{sma}), (\ref{smb}) and  (\ref{smc})
do not depend on the choice of the cutoff $a$.

\section{Yang-Mills at Finite Temperature}

\begin{figure}
\begin{center}
\includegraphics[width=8cm,clip]{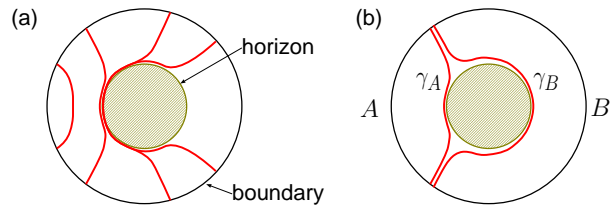}
\end{center}
\caption{ \label{fig: ads_blackhole.eps} (a) Minimal surfaces
$\gamma_A$ for various sizes of $A$. (b) $\gamma_A$ and $\gamma_B$
wrap the different parts of the horizon.}
\end{figure}

As the final example, we discuss the $\mathcal{N}=4$
super Yang-Mills theory on $\mathbb{R}^3$ at a finite
temperature $T$, which is dual to the AdS black hole
defined by the metric
\cite{Witten}
\be ds^2=R^2\left[\f{du^2}{hu^2}+u^2\left(-h
dt^2+dx_1^2+dx_2^2+dx_3^2\right)+d\Omega_5^2\right], \ee
where $h=1-u_0^4/u^4,\ u_0=\pi T$.
For the straight belt $A_{S}$,
the area is given by (putting the cut off $u\sim z^{-1}\sim a$)
\begin{eqnarray}
{\rm
Area}_{A_S}=2R^3L^2\int^{a^{-1}}_{u_*}
\f{du~ u^6}{\s{(u^4-u_0^4)(u^6-u_*^6)}}, \label{temparea}
\end{eqnarray}
where $u_*$ satisfies $ l/2=\int^{\infty}_{u_*}du
[(u^4-u_0^4)(u^6/u_*^6-1)]^{-1/2}. $ Eq.\ (\ref{temparea}) contains
the UV divergence $\sim a^{-2}$ as before. As in the analogous
computation of Wilson loops \cite{Wilsontemp}, we also expect a term
which is proportional to the area of $A$. Indeed, when $l$ is large
($u_*\sim u_0$) we find the constant term $\sim \pi^3R^3L^2lT^3$.
This leads to the finite part of  $S_A$ \be
S_{finite}\simeq\f{\pi^2}{2}N^2
T^3L^2l=\f{\pi^2N^2T^3}{2}\times\left({\rm Area~ of}~ A_{S}\right).
\label{vol}\ee We can regard this entropy as a part of the
Bekenstein-Hawking entropy of black 5-branes \cite{GKP}, which is
proportional to the area of horizon situated at $u=u_0$. Thus we can
interpret the part (\ref{vol}) as a thermal entropy contribution to
the total entanglement entropy at finite temperature. In our
gravitational description, this part arises because the minimal
surface wraps a part of the black hole horizon (Fig.\ \ref{fig:
ads_blackhole.eps} (a)). If we expand the size of $A$ until it
coincides with the total system (in the global coordinate),
$\gamma_A$ wraps the horizon completely and $S_A$ becomes equal to
the Bekenstein-Hawking entropy as expected. In a sense, the overall
normalization in Eq.\ (\ref{arealaw}) is fixed from Eq.\
(\ref{BHentropy}) once we consider the entanglement entropy at
finite temperature. Note that at finite temperature, $S_A = S_B$
does not hold generically. In the present situation, this occurs
because the surfaces $\gamma_A$ and $\gamma_B$ wrap two different
parts of the horizon (Fig.\ \ref{fig: ads_blackhole.eps} (b)).

As argued in \cite{MBH,Einhorn}, the AdS black hole can be dual to an entanglement
of two different CFTs at the two boundaries.  It is interesting to look at
the minimal surfaces
that connect them. As a specific limit, we may think the black hole
entropy is the same as the entanglement entropy of the CFTs  as the minimal surface
wrap the horizon.

\section{Conclusion}

In this paper we proposed a holographic description of the
entanglement entropy in quantum (conformal) field theories via
AdS$_{d+2}$/CFT$_{d+1}$ correspondence. This is summarized as the area
law relation (\ref{arealaw}). Based on our proposal we computed the
entanglement entropy for various systems,
e.g. 2D CFTs and the 4D
large-$N$ ${\mathcal{N}}=4$ super Yang-Mills theory. We
checked that in the lowest dimensional case ($d=1$), our formula
exactly coincides with the entropy computed directly from CFT.

In higher dimensions ($d\geq 2$),
a quantitative comparison is not easy because the gravity description
corresponds to the strongly coupled gauge theory whose entanglement
entropy is not known at present. Nevertheless, we found that
our computation for AdS$_5\times S^5$ reproduces the same functional
form of entanglement entropy as the one in the 4D free
${\mathcal{N}}=4$ super Yang-Mills. Their numerical coefficients
only differ by a factor $\sim \f{3}{2}$ in a rough estimation (assuming
that the gauge field contribution is the same as that of two
real scalar fields). This is similar to the well-known result of
the thermal entropy \cite{GKP}.

Finally we computed the entanglement entropy at finite temperature
employing AdS black hole geometry. In this case the minimal surface
wraps a fraction of the black hole horizon and this part is
responsible for the thermal contribution.


We would like to thank M. Einhorn, T. Harmark, G. Horowitz, Y. Hyakutake, D.
Marolf, T. Okuda, S.-J. Rey, M. Shigemori and Y. Sugawara for useful
discussions. This research was supported in part by the National
Science Foundation under Grant No.\ PHY99-07949.


\end{document}